\documentstyle[preprint,eqsecnum,aps,prb,amstex]{revtex}
\begin{document}
\title{Microphase separation in Pr$_{0.67}$Ca$_{0.33}$MnO$_3$ by small angle
neutron scattering}
\author{Ch. Simon, S. Mercone, N. Guiblin, C. Martin,}
\address{Laboratoire CRISMAT, Unit\'{e} Mixte de Recherches 6508, Institut des\\
Sciences de la Mati\`{e}re et du Rayonnement - Universit\'{e} de Caen, 6\\
Boulevard du\\
Mar\'{e}chal Juin, 14050 Caen Cedex, France.}
\author{A. Br\^{u}let, G. Andr\'{e},}
\address{Laboratoire L\'{e}on Brillouin, CE SACLAY, 91191 Gif/Yvette, France}
\date{\today}
\maketitle

\begin{abstract}
We have evidenced by small angle neutron scattering at low temperature the
coexistence of ferromagnetism (F) and antiferromagnetism (AF) in Pr$_{0.67}$%
Ca$_{0.33}$MnO$_3$. The results are compared to those obtained in Pr$_{0.80}$%
Ca$_{0.20}$MnO$_3$ and Pr$_{0.63}$Ca$_{0.37}$MnO$_3$, which are F and AF
respectively. Quantitative analysis shows that the small angle scattering is
not due to a mesoscopic mixing but to a nanoscopic electronic and magnetic
''red cabbage'' structure, in which the ferromagnetic phase exists in form
of thin layers in the AF matrix (stripes or 2D ''sheets'').
\end{abstract}


It was recently proposed that the ground state of manganites which displays
colossal magnetoresistive (CMR) properties \cite{review} could be an
electronic phase separation \cite{moreo,khomskii}. This is a very elegant
manner to interpret the CMR properties by percolation of a metallic
ferromagnetic phase in an insulating antiferromagnetic matrix. A small
change of the fraction or of the arrangement of the domains can induce
percolation. However, as pointed out by Radaelli et al. \cite{radaelli}, if
several works evidence phase separation, no clear results are published
about the size and shape of the objects. Some authors have observed small
(nanometric) ferromagnetic clusters by small angle neutron scattering \cite
{radaelli,deteresa,lynn} , while others found much larger phase separation
by high-resolution electron microscopy \cite{uehara}. Moreover, the width of
the magnetic Bragg peaks in neutron diffraction experiments is not
compatible with nanometric objects since the correlation length of the
magnetic order $\xi $ is usually more than 100 nm \cite{jirak}. The origin
of this phase separation can be different if it occurs on a nanoscopic or
mesosopic scale. To answer this question is one of the most important issues
in this research field.

The Pr$_{1-x}$Ca$_x$MnO$_3$ series is one of prime interest, because Pr and
Ca are about the same size and hence minimize the cationic size mismatch
effect. For compositions close to x=0.30, existence of the phase separation
is now well proved by magnetoresistance, magnetization and specific heat
studies\cite{hardy,deac}. Despite these numerous studies, the nature of this
phase separation is not solved at the present time. To summarize the
discussion, let us recall that the Pr$_{1-x}$Ca$_x$MnO$_3$ magnetic phase
diagram presents different states depending on the x value. For the higher Mn%
$^{3+}$ contents, the compounds are ferromagnetic at low temperature
(typically x=0.2), and for larger x values (typically x=0.4) they are
orbital ordered, antiferromagnetic CE like type. In between, the composition
x=0.33, studied here, shows a mixing of F and AF phases as shown by neutron
diffraction\cite{jirak}. From the width of the magnetic Bragg peaks, it is
clear that the order is mainly long range, typically more than 100 nm.
Paradoxically, small angle neutron scattering measurements exhibit a large
signal due to local inhomogeneities in the range of few nm \cite{radaelli}.
The conclusion of these previous experiments was that the sample is
inhomogeneous at a macroscopic scale, different zones being at the origin of
the long range order and the small angle neutron scattering signal.

Small angle neutron scattering is indeed a very powerful technique to study
the phase separation between a ferromagnet and an antiferromagnet since the
contrast between them is very large (the AF does not scatter at small
angle). It was first used in phase separated manganites by de Teresa et al. 
\cite{deteresa} and more recently by Radaelli\cite{radaelli}, but no
quantitative analysis of the scattered intensity, as we present here, was
performed.

Using the floating-zone method with feeding rods of nominal compositions Pr$%
_{0.6}$Ca$_{0.4}$MnO$_3$, Pr$_{0.7}$Ca$_{0.3}$MnO$_3$ and Pr$_{0.8}$Ca$%
_{0.2} $MnO$_3$, several-cm-long single crystals were grown in a mirror
furnace. Samples were cut out of the central part of these crystals and were
analyzed by EDS: their cationic compositions are homogeneous and were found
to be x=0.37, 0.33 and 0.20 respectively. Magnetization and transport
measurements were performed to check their quality. The samples were then
powdered in order to perform neutron diffraction. The powder diffraction
patterns were recorded in the G41 spectrometer in Orph\'{e}e reactor from
1.5K to 300K. The structures were refined using the Fullprof program, in the
classical Pbnm symmetry (a$_p\sqrt{2},$ a$_p\sqrt{2},$ 2a$_p$).

Classical behaviors were found:

x=0.37 is insulating and AF with T$_N$ at about 170K and with a structural
transition associated with the doubling of cell parameter at about 245K (T$%
_{CO}$). At low temperature, the spins are aligned along the c axis and the
magnetic order is that of the pseudo CE ordering \cite{jirak}.

x=0.20 presents a ferromagnetic phase transition with Tc at about 130K. The
spins are oriented along the a axis.

x=0.33 presents both a F component (similar to that of x=0.20) for half of
the spins (2$\mu _B$) and an AF one (similar to that of x=0.37) with 1/4 of
the spins. Tc is about 100K and T$_N$ about 115K.

Below 30K, there appears in the samples a contribution of the Pr moments
ordering which will not be discussed here.

Small angle neutron scattering were performed on PAXY spectrometer at the
Orph\'{e}e reactor. Two different experimental configurations were used: the
first is with a wavelength of 6 \AA\ and a sample-multidetector distance of
3.2m, allowing to study Q values between 10$^{-2}$ and 10$^{-1}$ \AA $^{-1}$
. The second one is with a wavelength of 4 \AA\ and a sample detector
distance of 1.2m, allowing to reach higher values up to 5 10$^{-1}$ \AA $%
^{-1}$. The sample was introduced in a cryostat with aluminum windows. In
order to subtract the background signal, an empty cell was measured. The
calibration of the spectrometer was performed with a Plexiglas sample
following the procedure given in reference\cite{neutronpa}. After
subtraction of the background, normalization by the Plexiglas sample, the
scattering function is presented in absolute units (cm$^{-1}$). We have
systematically neglected the inelastic spin wave corrections (the fact that
different neutron wavelengths give the same scattering curves supports this
assumption) and the magnetic form factor of the ions (this is valid at these
small Q values).

The small angle scattering in the x=0.37 sample presents the typical
features of an antiferromagnet (fig. 1) and Bragg diffraction exhibits nice
AF peaks which disappear at T$_N$=170K (inset of fig. 1). At small Q, the
small angle scattering is dominated by the granular structure of the powder:
it obeys to the classical Porod law and varies in Q$^{-4}$. In the
paramagnetic phase, it is due to the nuclear and non ordered magnetic
scattering. In the antiferromagnetic phase, only the nuclear contribution
remains, since the antiferromagnetic phase does not contribute to the small
angle scattering. The intensity of the Q$^{-4}$ signal can be calculated
from the grain size R by using

I(Q)= 4$\pi $ /R $\Delta \rho _{}^2$ Q$^{-4}$

where $\Delta \rho $ is the difference between the scattering length
densities ($\Delta \rho ^2$=3.2 10$^{21}$ cm$^{-4}$) between the two phases
(here the sample and the vacuum). Direct electron microscopy observations
show that the smallest grains are observed to be small flat grains of
thickness about R=3 $\mu $m. Then IQ$^4$ can be calculated to be 10$^{27}$ cm%
$^{-5}$ at room temperature. There is a perfect agreement with the value
measured by small angle neutron scattering of figure 1. At higher Q, the
signal measured is a paramagnetic contribution at high temperature, which
disappears completely in an antiferromagnetic phase. The low temperature
signal can be thus considered as the non magnetic signal even at high Q and
will be used as a background for the x=0.33 and the x=0.20 samples. Such an
assumption is verified since I(Q) at room temperature is exactly the same
for the three samples (not shown here).

The x=0.20 sample is ferromagnetic below 130K. The ferromagnetic component
of the diffraction peaks disappears at Tc (inset of fig. 2). The Q$^{-4}$
signal due to the granular structure of the sample is clearly observed above
and below Tc (fig. 2). In the neighborhood of Tc, the Lorentzian shape (Q$^2$%
+$\xi ^{-2}$)$^{-1}$ in which $\xi $ is the correlation length of the order
was previously used to describe this phase transition and is here adequate%
\cite{deteresa}. At low temperature, $\xi $ is very small (about few nm, not
shown in the figures).

The x=0.33 sample is a mixing of F and AF components below 110K. In the
upper inset of fig. 3, we have reported the temperature dependence of both F
and AF components on the Bragg peaks, showing this coexistence. In the lower
insets of fig. 3, it is shown that these Bragg peaks are narrow, close to
experimental resolution and certainly not typical of short range order. The
small angle scattering shown in fig. 3 is, at a first glance, very similar
to that of a ferromagnetic sample (compare to fig. 2). However, at low
temperature, the magnetic signal obtained by subtracting the signal of the
sample x=0.37 is very different for the two ferromagnetic samples as
presented in fig. 4. In addition to the Q$^{-4}$ Porod contribution\cite
{porod} due to the granular structure which exists in both samples, there is
a clear Q$^{-2}$ contribution in the high Q regime in the only x=0.33
sample. This Q dependence is that of infinite 2D sheets\cite{neutronpa}.
This analysis was introduced first in the soft matter field by Nallet et al.%
\cite{nallet}. In the present case, it corresponds to 2D ferromagnetic
sheets in antiferromagnetic matrix which plays the role of the vacuum. The
adjustable parameter in this model is the thickness of the 2D sheets ''t''.
For such a 2D object, the scattering function is I(Q)= 2$\pi $ $\phi (1-\phi
)$t $\Delta \rho _m^2$ Q$^{-2}\frac{(1-\cos Qt)}{(Qt)^2}$ which reduces in
the small Q limit to:

I(Q)= $\pi \phi (1-\phi )$ t $\Delta \rho _m^2$ Q$^{-2}$

$\Delta \rho _m$ is here proportional to the magnetization M: $\Delta \rho
_m $ =$\alpha M.$ M can be determined from the diffraction data or from the
small Q values as shown in the inset of fig. 3. (the proportional constant $%
\alpha $ is 0.27 10$^{-12}$ cm/$\mu _B$ divided by the unit volume of a Mn
cation 0.57 10$^{-22}$ cm$^3$, so $\Delta \rho _m^2$ =0.4 10$^{21}$ cm$^{-4}$%
). $\phi $ is the ferromagnetic fraction ($\phi $=0.5 here). Experimentally,
IQ$^2$=10$^{14}$ cm$^{-3}$. This drives to a parameter value t, which is
about 2.5 nm. Moreover, it can be stated that t does not vary significantly
with temperature up to the Curie phase transition. Close to the critical
temperature Tc, the shape of the scattering function changes, turning to the
Lorentzian function (Q$^2$+$\xi ^{-2}$)$^{-1}$ , classical of critical
fluctuations close to a phase transition\cite{deteresa}.

Note that spherical domains would have given a Q$^{-4}$ variation and linear
domains a Q$^{-1}$dependence\cite{neutronpa}. The Lorentzian function (Q$^2$+%
$\xi ^{-2}$)$^{-1}$ which was previously used to describe this phase
separation\cite{deteresa} is clearly not adequate below Tc. It would have
driven to very small $\xi $ values (smaller than 1\AA ), and has not
physical meaning. Note also that this Q$^{-2}$ contribution is very hard to
observe directly on the Bragg peaks since the convolution of this Q$^{-2}$
contribution with the lorentzian resolution of the spectrometer does not
broaden them, contrary to what is observed if $\xi $ is not zero. This
explains why there is no important broadening of the Bragg peaks, even with
this very large small angle neutron scattering (inset b of fig. 3). This
situation is also very different from that of uncorrelated 2D sheets which
should present asymmetrical Warren shapes of the Bragg peaks\cite{warren}.
An image of this microphase separation can be proposed in a ''red cabbage
illustration'' in which the ferromagnetic phase is the cabbage itself
whereas the antiferromagnetic part is the vacuum (inset of fig. 4). Such an
image can both explain the long range of the magnetic order and the very
large small angle signal with this specific Q dependence.

The phase separation occurs at about x=0.33, in between x=0.37 which is
purely AF and x=0.2 which is purely F. By looking to the magnetic properties
of the two phases and to the phase diagram, it can be suggested that a
charge transfer between the two phases can drive to a modulation of the
charge density (the cationic composition is not modified) from x=0.2 to
x=0.4. x=0.2 (1/5) is possibly an insulating ferromagnetic phase similar to
x=0.125 (1/8). Some recent theoretical work supports this point of view,
considering possible orbital ordering in x=1/8\cite{whystripes}. For the
antiferromagnetic phase, the CE structure of x=0.50 is not a possible
candidate because the antiferromagnetic phase of the x=0.33 is pseudo CE
type. Since it was shown that the x=0.37 sample presents about 2\% of
ferromagnetism\cite{hardy}, x=0.40 is probably typical of the charge density
of the F phase in x=0.33. Some more experimental work in needed to see in
the non powdered single crystal itself if the small angle scattering is
isotropic or not. It will be also very important to test the role of the
magnetic field on this striped phase. The nanoscopic structure proposed here
to interpret the apparent discrepancy between long range ferromagnetic order
and strong small angle scattering in x=0.33 sample was first proposed from
electrostatic arguments, due to the high Coulomb energy cost of the charge
separation\cite{moreo}. However, as pointed out by Khomskii et al.\cite
{whystripes}, electrostatic interaction is not enough to understand the
existence of the 2D sheets of the ''red cabbage'' system. Pure electrostatic
interactions should drive to charge and/or orbital ordering with a
superstructure (as it is observed in the x%
%TCIMACRO{\TEXTsymbol{>}}
%BeginExpansion
\mbox{$>$}%
%EndExpansion
0.5 part of the phase diagram), but not to 2D sheets structure. For x%
%TCIMACRO{\TEXTsymbol{<}}
%BeginExpansion
\mbox{$<$}%
%EndExpansion
0.5, elastic interactions would be the crucial ingredient to obtain such a
striped structure for the electronic phase separation.

In conclusion, we have shown that magnetic small angle neutron scattering is
particularly adapted to this study of phase separation in manganites. The
excellent contrast between  ferromagnetism and  antiferromagnetism provides
large signals. The analysis of the Q dependence associated with absolute
measurements allow to determine the shape and the size of the ferromagnetic
objects. In Pr$_{0.67}$Ca$_{0.33}$MnO$_3$, the prototype of electronic phase
separation, it appears a strong Q$^{-2}$ signal, characteristic of 2D
ferromagnetic stripes, which suggests a red cabbage structure. The thickness
of the stripes is about 2.5nm at 10K. This model also explains why there is
no significant broadening of the magnetic Bragg peaks. This technique offers
the possibility to investigate in details this new and important research
field and provides experimental data about the size and the shape of the
phase separation. 

Acknowledgments: We acknowledge sample preparation, characterizations and
many scientific discussions with M.\ Hervieu, L.\ Herv\'{e}, B.\ Raveau, A.\
Maignan, A.\ Wahl, D.\ Khomskii and V.\ Hardy. S. Mercone has been supported
by a Marie Curie fellowship of the European community program under contract
number HPMT2000-141.

Figure Captions

Figure 1: The small angle scattering functions of the x=0.37 sample (AF with
T$_N\simeq $170K) at two temperatures below and above T$_N$. The two
different Q ranges are evidenced by a small gap in the curves at about 0.1
\AA $^{-1}.$ Inset: the antiferromagnetic magnetization given by the
integration of the intensity of the antiferromagnetic 110 peak of powder
pattern obtained on G41.

Figure 2 : The small angle scattering functions of the x=0.20 sample (F with
T$_C\simeq $130 K) at three temperatures below, above and close to T$_C$.
Inset: the ferromagnetic magnetization given by the ferromagnetic 002 peak
intensity of powder pattern obtained on G41.

Figure 3: The small angle scattering functions of the x=0.33 sample (F+AF
with T$_N\simeq $T$_C$ $\simeq $ 110K) at three temperatures below, above
and close to the critical temperature. The antiferromagnetic and the
ferromagnetic magnetizations given by the intensities of the 110 and 002
peaks of powder pattern obtained on G41 are shown in the inset (a), as well
as the intensity in 10$^{-2}$ \AA $^{-1}$. In the inset (b), these typical
ferromagnetic or antiferromagnetic Bragg peaks show the absence of any
important broadening.

Figure 4 : The magnetic scattering function I$_m$(Q) for x=0.33 (F+AF with T$%
_N\simeq $T$_C$ $\simeq $ 110K) and x=0.20 (F with T$_C\simeq $130 K)at
T=10K, showing the Q$^{-4}$ and the Q$^{-2}$ behaviors. In the inset, a
schematic drawing of a cut of the nanoscopic electronic red cabbage model.

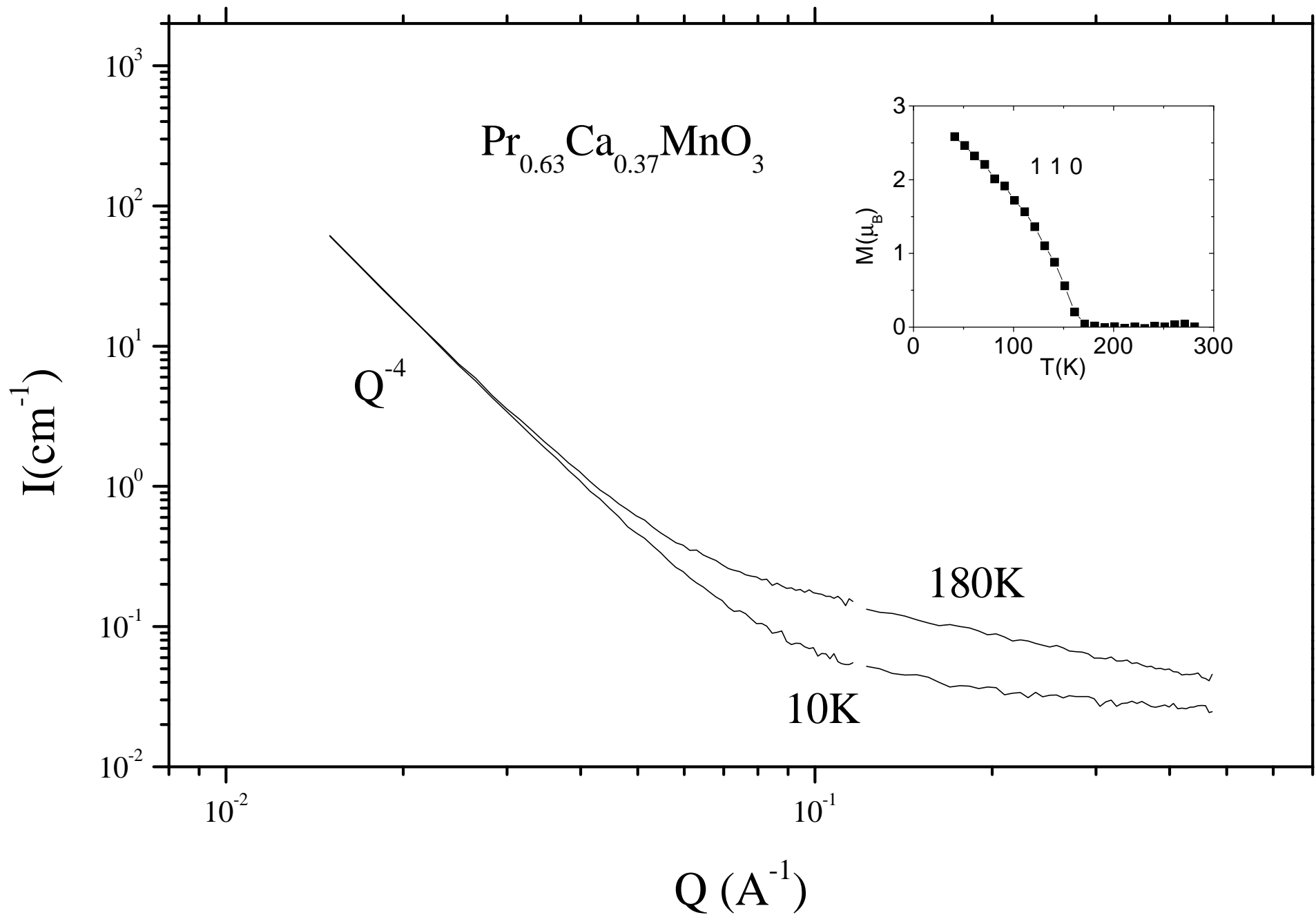
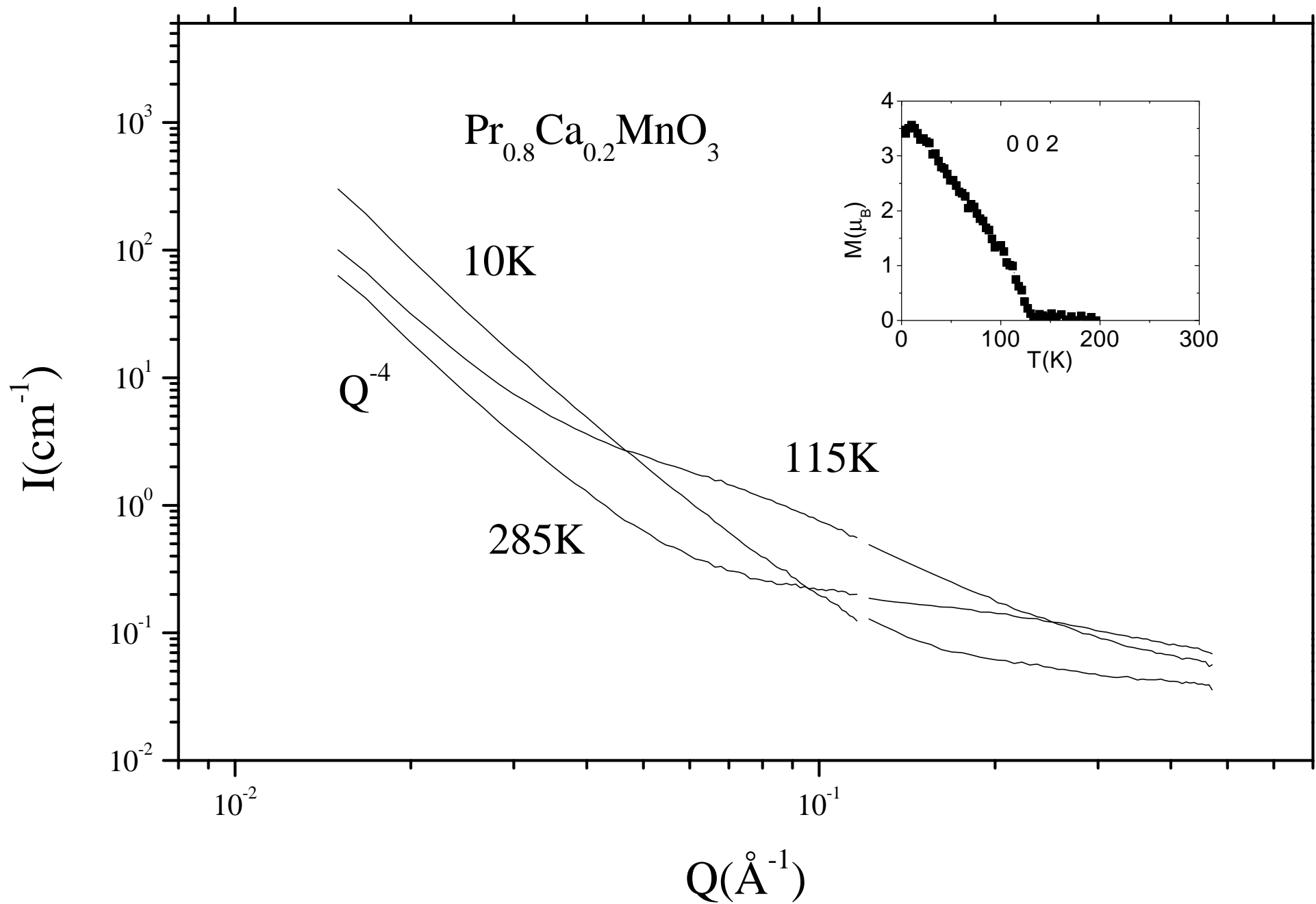
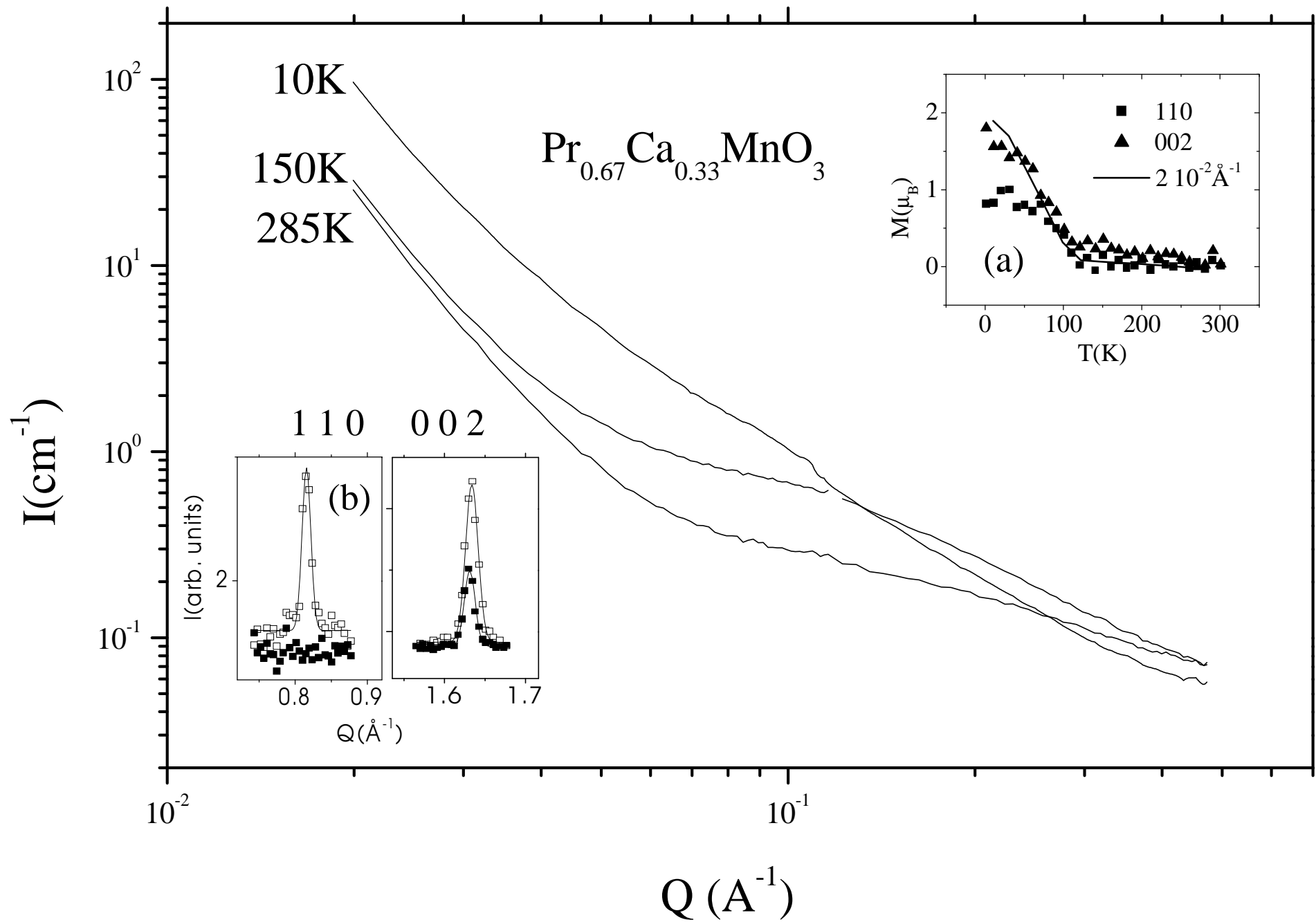
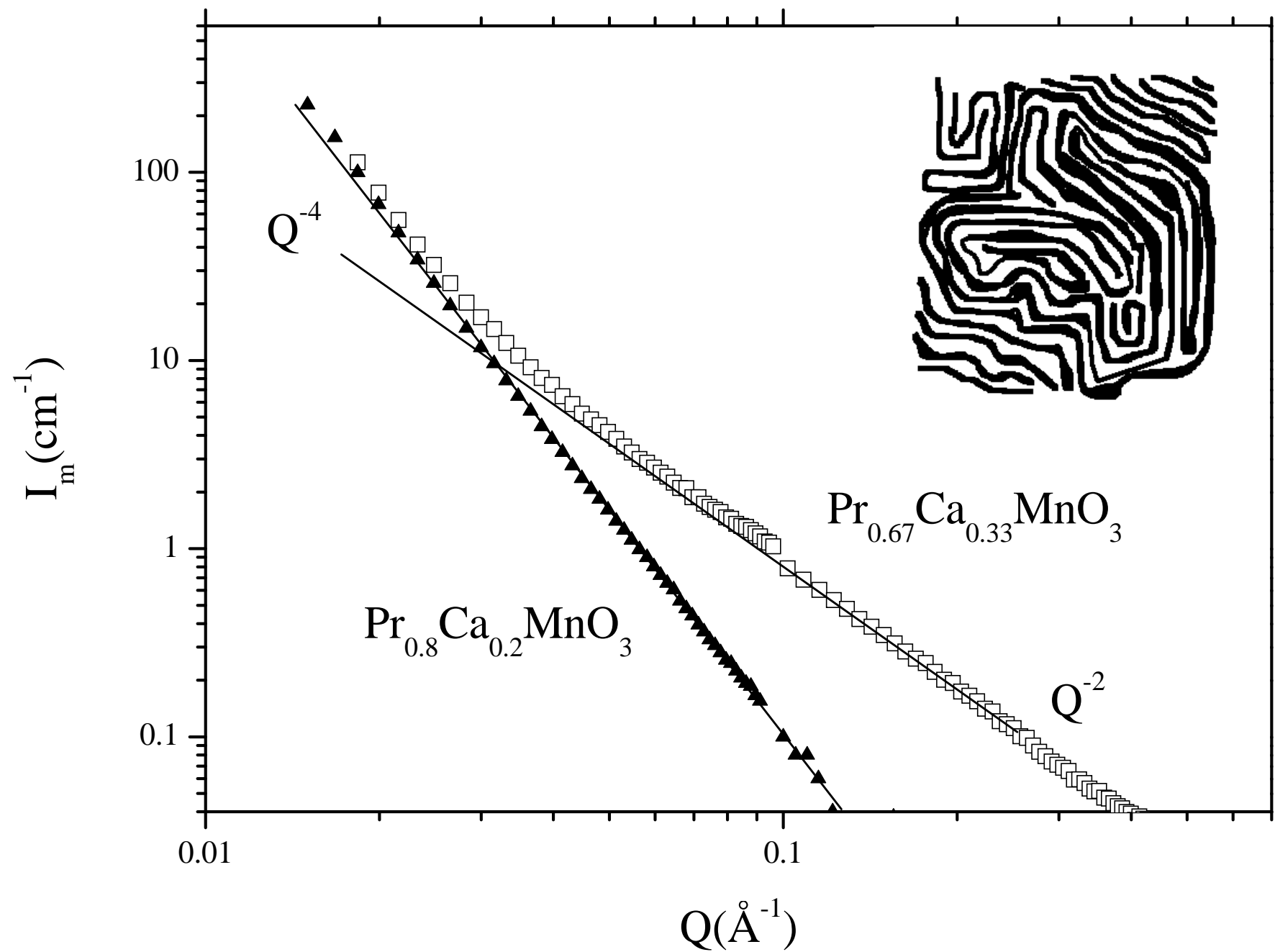
\end{document}